\documentclass[10pt]{article}
\usepackage[dvips]{color}
\usepackage{amsmath}
\usepackage{graphicx}
\begin{document}
\title{\bf Adiabatic expansion approximation solutions for the three-body problem}
\author{M.R. Pahlavani$^{a}$\thanks{e-mail:
m.pahlavani@umz.ac.ir}\hspace{1mm} and S.M.
Motevalli$^{a,b}$\thanks{e-mail:
motavali@umz.ac.ir}\\
$^a$ {\small {\em Department of Physics, Faculty of Science,
Mazandaran University,}}\\
{\small {\em P.O.Box 47415-416, Babolsar, Iran}}\\
$^b$ {\small {\em Department of Physics,Islamic Azad University,
Ayatollah Amoli Branch,}}\\
 {\small {\em P.O.Box 678, Amol,
Mazandaran, Iran}}} \maketitle
\begin{abstract} The motion of a muon in two centers coulomb field is one of the interesting problems of quantum mechanics. The adiabatic expansion method is powerful approach to study the muonic three-body system. In this investigation the three-body problem is studied for short-range interactions. Bound states and energy levels of this system were calculated
and compared with their Born-Oppenheimer method counterparts. The
obtained results are in good agreement with the previous
calculations.\\\\
{\bf M.S.C. 2000}: 36.10.-k, 21.45.+v
\end{abstract}
\section{Introduction}
A negative muon is a lepton of the second generation with  mass
number about  times heavier than that of electrons, and has a
finite lifetime of $\tau_{\mu}=2.197\mu sec$. Nuclear fusion
reactions can be catalyzed in a suitable fusion fuel by muons [1-3]. Energetic negative Muons, after stopping in a
hydrogen isotope $H_{2}/D_{2}/T_{2}$ mixture, fuse with different
hydrogen isotopes to form  the muonic atoms $p\mu$, $d\mu$ and
$t\mu$ in excited states. The size of a muonic atom is about $209$
times smaller than a normal atom. After a sequence of cascade
transitions lasting about $10^{-11}sec$ at liquid hydrogen density
($L.H.D \approx 4.25 \times 10^{22}atoms/cm^{3}$), muonic atoms
are formed in the ground $1S$-state with kinetic energy in the
range of $10^{-3}-10^{2}eV$, depending on the temperature of the
target and the prehistory of cascade transitions[4,5].
Muonic molecular formation may take place in collisions of the
muonic
atom in its ground state with ordinary hydrogen molecules.\\
The crucial process is the resonant formation of the muonic molecule
$dt\mu$ into the loosely bound rotational-vibrational state  $J=1$,
$\nu=1$ ($10^{-8}-10^{-10}sec$). This state quickly de-excites to
$J=0$  levels, from which the two nuclei fuse ($\approx 10^{-12}$).
Usually the muon is free to start the next cycle, but occasionally
(with probability  $\omega_{s}\simeq 0.555\%$ in a D-T target at
density $\phi=1.2$ $L.H.D$ [6]) it is removed from the active cycle
by "sticking" to the helium nucleus. Resonant formation of the
$dd\mu$ molecule at very low temperatures was observed in solid and
liquid $D_{2}$ targets [7,8]. Energy levels of muonic molecules in
muon catalyzed fusion ($\mu CF$) systems have been extensively
studied with various calculational methods [9-12]. The solution of
the non-relativistic Schrodinger equation for the three-body system
was found using two different methods: by: the adiabatic expansion
method and Born-Oppenheimer method. The Born-Oppenheimer approach
assumes the nuclei to be infinitely heavy with respect to the
negatively charged particle. It should be kept in mind that the
following Born-Oppenheimer approach is the simplest solution to the
three-body coulomb system. This approach is expected to be a poor
approximation for calculations of muonic molecule eigenvalues. In
the present paper, we calculate binding energies of the bound states
of the $dd\mu$ muonic molecule using the adiabatic expansion method.
\section{Theoretical calculations}
The exact Hamiltonian that describes muonic three-body system can be shown by following relation:\\
\begin{equation}
H=-\frac{1}{2m_{1}}\nabla_{R_{1}}^{2}-\frac{1}{2m_{2}}\nabla_{R_{2}}^{2}-\frac{1}{2m_{\mu}}\nabla_{R_{\mu}}^{2}-\frac{z_{1}z_{\mu}}{|\textbf{r}_{\mu}-\textbf{R}_{1}|}-\frac{z_{2}z_{\mu}}{|\textbf{r}_{\mu}-\textbf{R}_{2}|}+\frac{z_{1}z_{2}}{|\textbf{R}_{1}-\textbf{R}_{2}|}\\
\end{equation}
where 1 and 2 denote the two nuclei, their position is given by
$R_{1}$ and $R_{2}$, and the muon coordinate is $r_{\mu}$. The
center of mass coordinate $R_{CM}$ is given by
\begin{equation}
\textbf{R}_{CM}=\frac{m_{1}\textbf{R}_{1}+m_{2}\textbf{R}_{2}+m_{\mu}\textbf{r}_{\mu}}{m_{1}+m_{2}+m_{\mu}}\\
\end{equation}
It is convenient to define Jacobi coordinate $r$  and $R$ as
follow:
\begin{eqnarray}
&&\textbf{r}=\textbf{r}_{\mu}-\frac{\textbf{R}_{1}+\textbf{R}_{2}}{2}\\
&&\textbf{R}=\textbf{R}_{2}-\textbf{R}_{1}\nonumber
\end{eqnarray}
where  $R$ is the internuclear coordinate and $r$ is the muon
coordinate to midpoint between the two nuclei. In these
coordinates ($R,r$), the Hamiltonian denoted by equation (1) is
change to the following operator:
\begin{equation}
H=-\frac{1}{2M_{T}}\nabla_{R_{CM}}^{2}-\frac{1}{2M}\left(\nabla_{R}+\frac{\lambda}{2}\nabla_{r}\right)^{2}-\frac{1}{2m}\nabla_{r}^{2}-\frac{z_{1}z_{\mu}}{|\textbf{r}_{\mu}-\textbf{R}_{1}|}-\frac{z_{2}z_{\mu}}{|\textbf{r}_{\mu}-\textbf{R}_{2}|}+\frac{z_{1}z_{2}}{|\textbf{R}_{1}-\textbf{R}_{2}|}\\\\
\end{equation}
where
\begin{eqnarray}
&&M_{T}=m_{1}+m_{2}+m_{\mu}\\
&&\frac{1}{M}=\frac{1}{m_{1}}+\frac{1}{m_{2}}\nonumber\\
&& \frac{1}{m}=\frac{1}{m_{1}+m_{2}}+\frac{1}{m_{\mu}}\nonumber\\
&&\lambda=\frac{m_{1}-m_{2}}{m_{1}+m_{2}}\nonumber
\end{eqnarray}
After separation of variables, the non-relativistic Hamiltonian in units of $e=\hbar=m_{\mu}=1$, can be given by
\begin{equation}
H(\hat{o},R)=-\frac{1}{2M}\left(\nabla_{R}+\frac{\lambda}{2}\nabla_{r}\right)^{2}-H_{1}+\frac{z_{1}z_{2}}{R}
\end{equation}
where
\begin{equation}
H_{1}=-\frac{1}{2m}\nabla_{r}^{2}-\frac{z_{1}}{r_{1}}-\frac{z_{2}}{r_{2}}
\end{equation}
where $\hat{o}$ represent the  five dimensional variable. We use the set $\hat{o}=(\Theta, \Phi, \xi, \eta, \varphi)$
where $(\Theta, \Phi, \varphi)$ define the Euler rotation specifying the body-fixed frame with its unit vectors to coincide
with the principal axes of the inertia tensor of a three-body system.
The hyperspheroidal coordinates $\xi$ and $\eta$ are easily expressed by the muon-nucleus
distances $r_{1}$, $r_{2}$ and the internuclear distance $R$,
\begin{eqnarray}
&&\xi=\frac{r_{1}+r_{2}}{R},~~(1\leq\xi<\infty)\\
&&\eta=\frac{r_{1}-r_{2}}{R},~~(-1\leq\eta\leq1) \nonumber
\end{eqnarray}
The three-body Hamiltonian $(6)$ commutes with the total angular
momentum operator for the three particle system, $\textbf{J}$, its
projection on z-axis, $\textbf{J$_{z}$}$, and the total parity
operator, $\textbf{P}(\textbf{R} \longrightarrow \textbf{-R},
\textbf{r} \longrightarrow \textbf{-r})$. Eigenfunctions of the
Hamiltonian in the total angular momentum representation reads:
\begin{equation}
\Psi_{J,M}^{p}(\textbf{R},\textbf{r})=\sum_{m=-J}^{J}F_{m}^{Jp}(R,\xi,\eta)D_{Mm}^{Jp}(\Phi,\Theta,\varphi)
\end{equation}
Adiabatic expansion of radial function, $F_{m}^{Jp}(R,\xi,\eta)$
is usually written in the form:
\begin{equation}
F_{m}^{Jp}(R,\xi,\eta)=\sum_{N=1}^{\infty}\sum_{l=0}^{N-1}\psi_{Nlm}(R;\xi,\eta)\chi_{Nlm}^{Jp}(R)R^{-1}+\sum_{l=0}^{\infty}\int_{0}^{\infty}dk\psi_{klm}(R;\xi,\eta)\chi_{klm}^{Jp}(R)R^{-1}
\end{equation}
where  $\chi_{i}^{Jp}(R)$ describe relative motion of the nuclei.
Let us consider the Wigner function,
$D_{Mm}^{Jp}(\Phi,\Theta,\varphi)$ which is the eigenstates of
$\textbf{J$^{2}$}$, $\textbf{J$_{z}$}$ and \textbf{R}.\textbf{J}/R
with the eigenvlues $J(J+1)$, $M$ and $m$ [13]. It can be
transformed under the inversion as follow:
\begin{equation}
\textbf{P}D_{Mm}^{J}(\Phi,\Theta,\varphi)=D_{Mm}^{J}(\Phi+\pi,\pi-\Theta,\pi-\varphi)=(-1)^{J-m}D_{M,-m}^{J}(\Phi,\Theta,\varphi)
\end{equation}
If  $m\neq0$, the resultant Wigner functions would be different, and the angular functions consist both even and odd combinations.
It is convenient to specify these combinations as follows:
\begin{equation}
D_{Mm}^{Jp}(\Phi,\Theta,\varphi)=\frac{\sqrt{2J+1}}{4\pi}\left[(-1)^{m}D_{Mm}^{J}(\Phi,\Theta,\varphi)+p(-1)^{J}D_{M,-m}^{J}(\Phi,\Theta,\varphi)\right]
\end{equation}
$P=\pm(-1)^{J}$ is the eigenvalue of the parity operator:
\begin{equation}
\textbf{P}D_{Mm}^{Jp}=pD_{Mm}^{Jp}
\end{equation}
The functions presented in equation (12)(in bracket) are satisfy the following orthonormality condition:
\begin{equation}
\int_{0}^{\pi}sin\Theta d\Theta
\int_{0}^{2\pi}d\Phi\int_{0}^{2\pi}
d\varphi\left[D_{Mm}^{Jp}(\Phi,\Theta,\varphi)\right]^{*}D_{M^{'}m^{'}}^{J^{'}p^{'}}(\Phi,\Theta,\varphi)=\delta_{JJ^{'}}\delta_{pp^{'}}\delta_{MM^{'}}\delta_{mm^{'}}
\end{equation}
If  $m=0$, both the Wigner functions in (11) are reduced to the ordinary spherical function $Y_{JM}(\Theta,\Phi)$ so that the dependence of $\varphi$ disappears and
the angular functions satisfying the conditions (13) and (14) are:
\begin{equation}
D_{M,m=0}^{Jp}(\Phi,\Theta,\varphi)=\frac{Y_{JM}(\Theta,\Phi)}{\sqrt{2\pi}}
\end{equation}
In this case the parity is unambiguously specified by the quantum
number $J$: $p=+(-1)^{J}$. So, our basis functions have the
following structure:
\begin{equation}
\Psi_{Mjm}^{Jp}(R,\Theta,\Phi,\xi,\eta,\varphi)=D_{M,m}^{Jp}(\Phi,\Theta,\varphi)\psi_{jm}(\xi,\eta;R)\frac{\chi_{jm}^{Jp}(R)}{R}
\end{equation}
The wave functions
$\Psi_{Mjm}^{Jp}(R,\Theta,\Phi,\xi,\eta,\varphi)$ describing
reactions $h\mu+h$, $h=(p,d,t)$ can be decomposed over the
solutions $\psi_{jm}(\xi,\eta;R)$ of the Coulomb two-center
problem. $\psi_{jm}(\xi,\eta;R)$ is the complete set of solutions
of the Coulomb two-center problem, therefore
\begin{equation}
H_{1}\psi_{i}(\xi,\eta;R)F(\varphi)=E_{i}(R)\psi_{i}(\xi,\eta;R)F(\varphi)
\end{equation}
describing the muon motion around fixed nuclei separated by a
distance $R$. $E_{i}(R)$ is the energy of a muon in the state $i$
as a function of $R$. Here we show how to separate the variables
through the use of the ellipsoidal (or, prolate spheroidal)
coordinates
\begin{eqnarray}
&&x=\frac{R}{2}\sqrt{(\xi^{2}-1)(1-\eta^{2})}cos\varphi\\
&&y=\frac{R}{2}\sqrt{(\xi^{2}-1)(1-\eta^{2})}sin\varphi \nonumber\\
&&z=\frac{R}{2}\xi\eta \nonumber
\end{eqnarray}
Note that the coordinates $\xi$, $\eta$ and $\varphi$ are orthogonal, and we have the first fundamental form
\begin{equation}
ds^{2}=dx^{2}+dy^{2}+dz^{2}=h_{\xi}^{2}d\xi^{2}+h_{\eta}^{2}d\eta^{2}+h_{\varphi}^{2}d\varphi^{2}
\end{equation}
where
\begin{eqnarray}
&&h_{\xi}^{2}=\left(\frac{\partial x}{\partial\xi}\right)^{2}+\left(\frac{\partial y}{\partial\xi}\right)^{2}+\left(\frac{\partial z}{\partial\xi}\right)^{2}=\frac{R^{2}}{4}\left(\frac{1-\eta^{2}}{\xi^{2}-1}\right)\\
&&h_{\eta}^{2}=\left(\frac{\partial x}{\partial\eta}\right)^{2}+\left(\frac{\partial y}{\partial\eta}\right)^{2}+\left(\frac{\partial z}{\partial\eta}\right)^{2}=\frac{R^{2}}{4}\left(\frac{\xi^{2}-1}{1-\eta^{2}}\right) \nonumber\\
&&h_{\varphi}^{2}=\left(\frac{\partial x}{\partial\varphi}\right)^{2}+\left(\frac{\partial y}{\partial\varphi}\right)^{2}+\left(\frac{\partial z}{\partial\varphi}\right)^{2}=\frac{R^{2}}{4}\left(\xi^{2}-1\right)\left(1-\eta^{2}\right)
\nonumber
\end{eqnarray}
Thus
\begin{eqnarray}
\nabla^{2}=\frac{1}{h_{\xi}h_{\eta}h_{\varphi}}\left[\frac{\partial}{\partial\xi}\left(\frac{h_{\eta}h_{\varphi}}{h_{\xi}}\frac{\partial}{\partial\xi}\right)+\frac{\partial}{\partial\eta}\left(\frac{h_{\xi}h_{\varphi}}{h_{\eta}}\frac{\partial}{\partial\eta}\right)+\frac{\partial}{\partial\varphi}\left(\frac{h_{\xi}h_{\eta}}{h_{\varphi}}\frac{\partial}{\partial\varphi}\right)\right]\\
=\frac{4}{R^{2}(\xi^{2}-\eta^{2})}\left\{\frac{\partial}{\partial\xi}\left[(\xi^{2}-1)\frac{\partial}{\partial\xi}\right]+\frac{\partial}{\partial\eta}\left[(1-\eta^{2})\frac{\partial}{\partial\eta}\right]+\frac{\xi^{2}-\eta^{2}}{(\xi^{2}-1)(1-\eta^{2})}\frac{\partial^{2}}{\partial\varphi^{2}}\right\}
\nonumber
\end{eqnarray}
Note that through the coordinate transformation (18), we have
\begin{eqnarray}
r_{1}=\frac{R}{2}(\xi+\eta)\\
r_{2}=\frac{R}{2}(\xi-\eta) \nonumber
\end{eqnarray}
Writing the wavefunction as $\psi_{i}(\xi,\eta;R)F(\varphi)=G(\xi)H(\eta)F(\varphi)$ and changing the variable to spheroidal coordinates,
equation (17) can be separated into following three one-dimensional equations:
\begin{equation}
\frac{d^{2}F(\varphi)}{d\varphi^{2}}+m^{2}F(\varphi)=0
\end{equation}
\begin{equation}
\frac{d}{d\xi}\left[(\xi^{2}-1)\frac{dG(\xi)}{d\xi}\right]+\left(-A+\alpha
q\xi-q^{2}\xi^{2}-\frac{m^{2}}{\xi^{2}-1}\right)G(\xi)=0
\end{equation}
\begin{equation}
\frac{d}{d\eta}\left[(1-\eta^{2})\frac{dH(\eta)}{d\eta}\right]+\left(+A-\beta
q\eta+q^{2}\eta^{2}-\frac{m^{2}}{1-\eta^{2}}\right)H(\eta)=0
\end{equation}
where
\begin{eqnarray}
\alpha=\frac{R}{q}(z_{1}+z_{2})\\
\beta=\frac{R}{q}(z_{1}-z_{2}) \nonumber
\end{eqnarray}
Note that $A$ and $q$ are unknown parameters and should be
obtained from (24) and (25) as eigenvalues of the coupled system.
Once $A$ and $q$ are obtained, then $E$ can be obtained from
$q^{2}=-R^{2}E/2$. By substitution of expression (16) into the
Schrodinger equation with Hamiltonian (6) and after averaging over
spherical angles $(\Theta,\Phi)$ and the muon  state, one obtains
the radial equation
\begin{equation}
\frac{1}{2M}\frac{d^{2}}{dR^{2}}\chi_{i}^{J}(R)+\left[\varepsilon-V_{B-O}(R)-U_{i}(R)-\frac{J(J+1)}{2MR^{2}}\right]\chi_{i}^{J}(R)=0
\end{equation}
where $\varepsilon=E-E_{i}(\infty)$ is the collision energy and
$E$ is the total energy of the system and $E_{i}(\infty)$ is the
ground state energy of muonic atom.
$V_{B-O}(R)=E-E(\infty)+\frac{z_{1}z_{2}}{R}$ is the potential
corresponding to the Born-Oppenheimer (B-O) approximation and
$U_{i}(R)=\langle
i|\left(H-H_{1}-\frac{z_{1}z_{2}}{R}\right)|i\rangle$ is the
adiabatic correction. The adiabatic potential $V_{Ad}(R)$ is:
\begin{equation}
V_{Ad}(R)=V_{B-O}(R)+U_{i}(R)
\end{equation}
\begin{tabular*}{2cm}{cc}
\hspace{1cm}\includegraphics[scale=1]{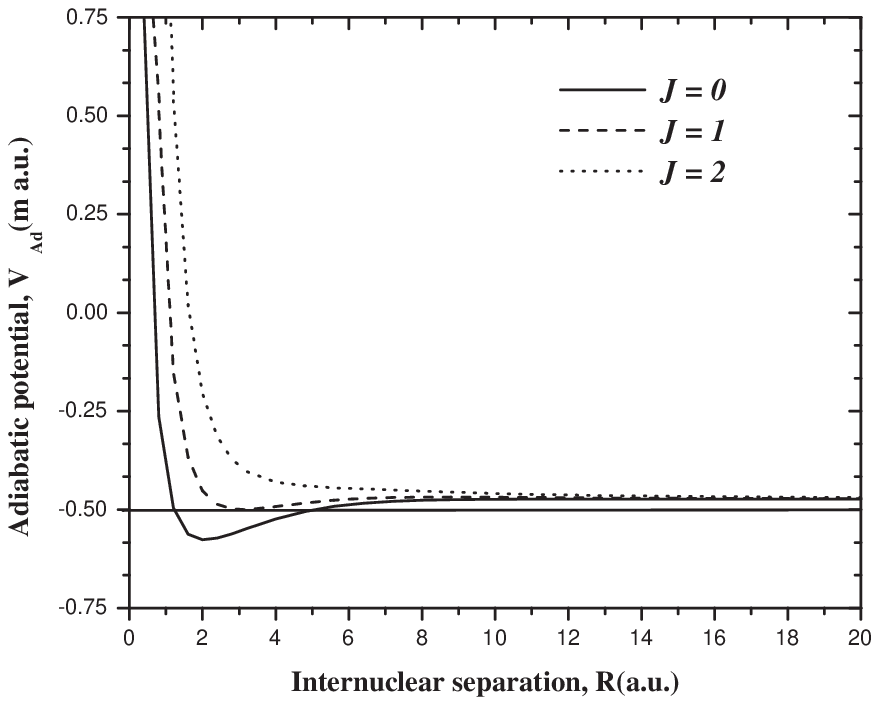}\\
\end{tabular*}\\
Figure 1: Adiabatic potential curves, $V_{Ad}$(J = 0, 1 and 2),
corresponding to $d-d-\mu$ system.\\\\
The adiabatic potential $V_{Ad}(R)$ for the $(dd\mu)$ muonic
molecule is calculated in the adiabatic expansion method. The
adiabatic potential curves and qualitatively similar for each of
muonic molecules and are displayed for $(dd\mu)$ in Figure 1.
Results of the calculations of binding energies of the bound states
$(J,\nu)$ of the $(dd\mu)$ muonic molecule are compared with the
results of the other methods used in Refs. [14,15] and are given in
Table 1.\\
\begin{table}
\caption{\label{tabone}Binding energies $(eV)$ of the states
$(J,\nu)$ for the $(dd\mu)$ muonic molecule.}
\center\begin{tabular}{|c|c|c|c|c|c} \hline
$(J,\nu)$ &Ad(This work)&Ref. [14]&Ref. [15]\\
\hline\hline
$(0,0)$&325.06&325.0735& 325.070 540  \\
\hline
$(0,1)$ & 35.79&35.8436&35.844 227   \\
\hline
$(1,0)$ &226.62&226.6815&226.679 792 \\
\hline
$(1,1)$ & 1.73&1.97475&1.974 985 \\
\hline
$(2,0)$ & 86.20&86.4936&--- \\
\hline
\end{tabular}
\end{table}
\section{conclusion}
In this paper, we have presented the calculation of the binding energies of the symmetric muonic molecular ion. Early calculations were unable to demonstrate the existence of the crucial weakly bound states.
For example, in the Born-Oppenheimer (fixed nuclei) approximation the state is much too bound, but if adiabatic corrections are included it is not bound at all. The results are obtained by the three body Hamiltonian in
the adiabatic expansion method.
The calculated binding energies are in good agreement with the previous calculations by other authors
using different methods.

\end{document}